\documentclass[12pt]{iopart}
\usepackage{graphicx}

\newcommand{\ud}{\mathrm{d}}

\begin{document}

\title{Statistical hadronization of charm: from FAIR to the LHC}

\author{A Andronic$^1$, P Braun-Munzinger$^{1,2}$, K Redlich$^{3,2}$, J Stachel$^4$}
\address{$^1$ GSI, Darmstadt, $^2$ Technical University, Darmstadt,
$^3$ Institute of Theoretical Physics, University of Wroc\l aw,
$^4$ Physikalisches Institut der Universit\"at Heidelberg}

%\ead{A.Andronic@gsi.de}

\begin{abstract}
We discuss the production of charmonium in nuclear collisions within the 
framework of the statistical hadronization model. 
We demonstrate that the model reproduces very well the availble data
at RHIC. We provide predictions for the LHC energy where, dependently on the
charm production cross section, a dramatically different behaviour of 
charmonium production as a function of centrality might be expected.
We extend our predictions for charm production towards the threshold energies,
where charm is expected to be measured at the future FAIR facility.
\end{abstract}


%\maketitle

%\section{Introduction}
Charmonium production is considered, since the original proposal more than 20
years ago about its suppression in a Quark-Gluon Plasma (QGP)
\cite{satz}, an important probe to determine the degree of deconfinement
reached in the fireball produced in ultra-relativistic nucleus-nucleus
collisions. 
In the original scenario of J/$\psi$ suppression via Debye screening
\cite{satz} it is assumed that the charmonia are rapidly formed in initial
hard collisions but are subsequently destroyed in the QGP (see an update
in ref. \cite{satz2}).

In recent publications \cite{aa2} we have demonstrated that the data 
on J/$\psi$ and $\psi'$ production in nucleus-nucleus collisions at the 
SPS ($\sqrt{s_{NN}} \approx 17$ GeV) and RHIC ($\sqrt{s_{NN}}$=200 GeV) energies 
can be well described within the statistical hadronization model proposed 
in \cite{pbm1} and have provided predictions for the LHC energy 
($\sqrt{s_{NN}}$=5.5 TeV) and for energies close to threshold 
($\sqrt{s_{NN}} \approx 6$ GeV).

%\section{The statistical hadronization model} 

In our statistical hadronization model (SHM) \cite{pbm1,aa1,aa2} we assume
that the charm quarks are produced in primary hard collisions and that their 
total number stays constant until hadronization at chemical freeze-out.  
Another important element is thermal equilibration in the QGP, at least 
near the critical temperature, $T_c$. 
Measurements of elliptic flow of J/$\psi$ will be a crucial check of this 
scenario.
%Charmonium production in the nuclear corona is taken into account \cite{aa2}.

The model has the following input parameters:
i) the charm production cross section in pp collisions, taken either from NLO 
pQCD calculations \cite{cac,rv1} or from experiment \cite{cc1};
ii) characteristics at chemical freeze-out: temperature, $T$, 
baryochemical potential, $\mu_b$, and volume corresponding to one unit 
of rapidity $V_{\Delta y=1}$, extracted from thermal fits of non-charmed 
hadrons \cite{aat}. 

%\section{Comparison to data and predictions}

\begin{figure}[htb]
\vspace{-.5cm}
%\begin{tabular}{cc} \begin{minipage}{.69\textwidth}
%charm#jpsi-raa2 ym=0.84
\centering\includegraphics[width=.84\textwidth]{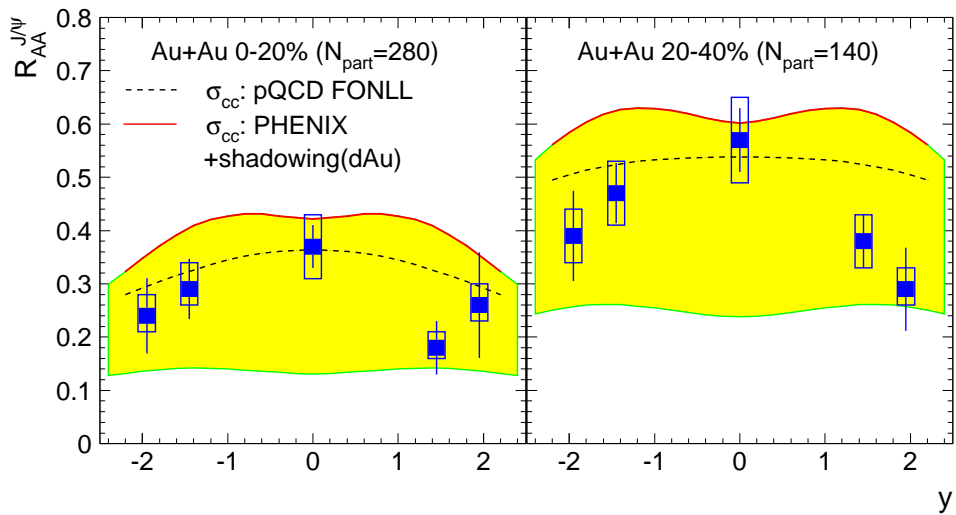}
%\end{minipage}  & \begin{minipage}{.29\textwidth}
\vspace{-.5cm}
  \caption{Rapidity dependence of $R_{AA}^{J/\psi}$ for two centrality 
classes. The data from the PHENIX experiment \cite{phe1} (symbols with errors)
are compared to calculations (lines, see text). The shaded area corresponds to 
calculations for the lower limit of the charm cross section as measured by 
PHENIX \cite{cc1}, with our shadowing scenario.}
\label{aa_fig1}
%\end{minipage}\end{tabular}
\end{figure}

In Fig.~\ref{aa_fig1} we present the rapidity dependence of the nuclear 
modification factor $R_{AA}^{J/\psi}$.
While earlier \cite{aa2} we have compared data \cite{phe1} to our model 
predictions for the pQCD charm production cross section \cite{cac} 
(dashed line in Fig.~\ref{aa_fig1}), we use here as alternative the charm 
cross section as measured by PHENIX in pp collisions \cite{cc1} and consider 
in addition shadowing for Au-Au collsions as extracted from recent 
dAu data \cite{phe2} (assuming teh deviation of $R_{dAu}^{J/\psi}$ from unity 
is due to shadowing).
Also in this case, our model describes the observed suppression and its 
rapidity dependence.  
The maximum of $R_{AA}^{J/\psi}$ at midrapidity is in our model due to the
enhanced generation of charmonium around mid-rapidity, determined by the
rapidity dependence of the charm production cross section. In this sense, 
the above result constitutes strong evidence for the statistical
generation of J/$\psi$ at chemical freeze-out.

\begin{figure}[htb]
\begin{tabular}{cc}
\begin{minipage}{.49\textwidth}
\centering\includegraphics[width=.94\textwidth]{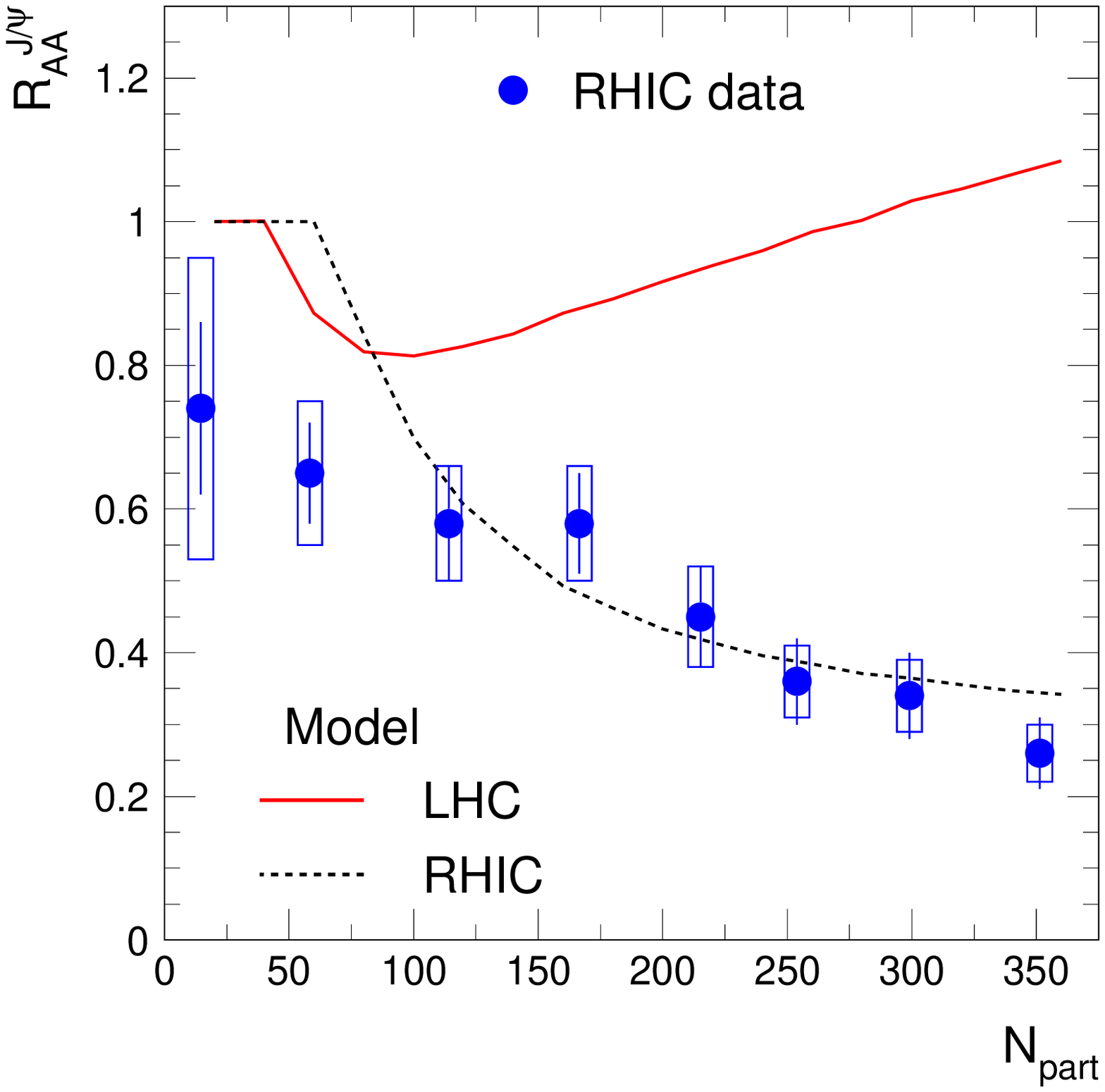}
\end{minipage}  & \begin{minipage}{.49\textwidth}
\centering\includegraphics[width=.94\textwidth]{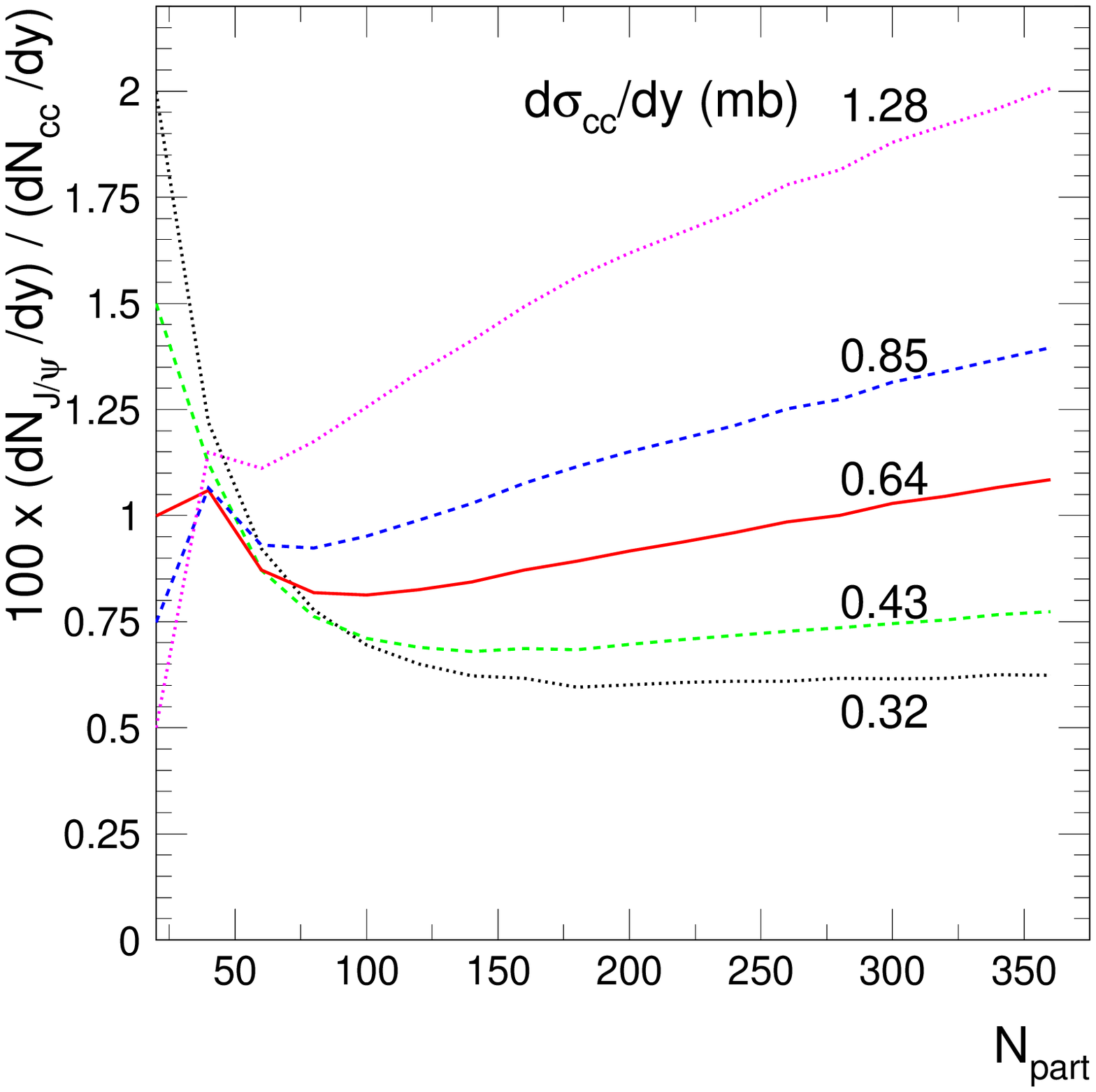}
\end{minipage}\end{tabular}
\vspace{-.5cm}
\caption{Centrality dependence of $R_{AA}^{J/\psi}$ for RHIC and LHC energies
(left panel) and of the $J/\psi$ rapidity density at LHC relative to the number 
of initially produced $c\bar{c}$ pairs (right panel, curves labelled by the
$\ud\sigma_{c\bar{c}}/\ud y$) at midrapidity.}
\label{aa_fig2}
\end{figure}

The centrality dependence of $R_{AA}^{J/\psi}$ at midrapidity is shown in
the left panel of Fig.~\ref{aa_fig2}.  Our calculations approach the value 
in pp collisions around $N_{part}$=50, which corresponds to an assumed minimal 
volume for the creation of QGP of 400 fm$^3$ \cite{aa2}.  
The model reproduces very well the decreasing trend versus centrality seen 
in the RHIC data \cite{phe1}.  
Note that, in our model, the centrality dependence of the nuclear modification 
factor arises entirely as a consequence of the still rather moderate rapidity 
density of initially produced charm quark pairs at RHIC ($\ud N_{c\bar{c}}/\ud y$=1.6
for central collisions, using the FONLL charm production cross section \cite{cac}).
At the much higher LHC energy the charm production cross section (including 
shadowing in PbPb collisions \cite{rv1}) is expected to be about an order 
of magnitude larger.  
As a result, the opposite trend as a function of centrality is predicted, 
with $R_{AA}^{J/\psi}$ exceeding unity for central collisions.  A significantly 
larger enhancement of about a factor of 2 is obtained if the charm production 
cross section is two times larger than presently assumed, as seen in the right panel
of Fig.~\ref{aa_fig2}, where we show the $J/\psi$ production relative to the 
number of initially produced $c\bar{c}$ pairs.

\begin{figure}[htb]
\begin{tabular}{cc}
\begin{minipage}{.49\textwidth}
\centering\includegraphics[width=.94\textwidth]{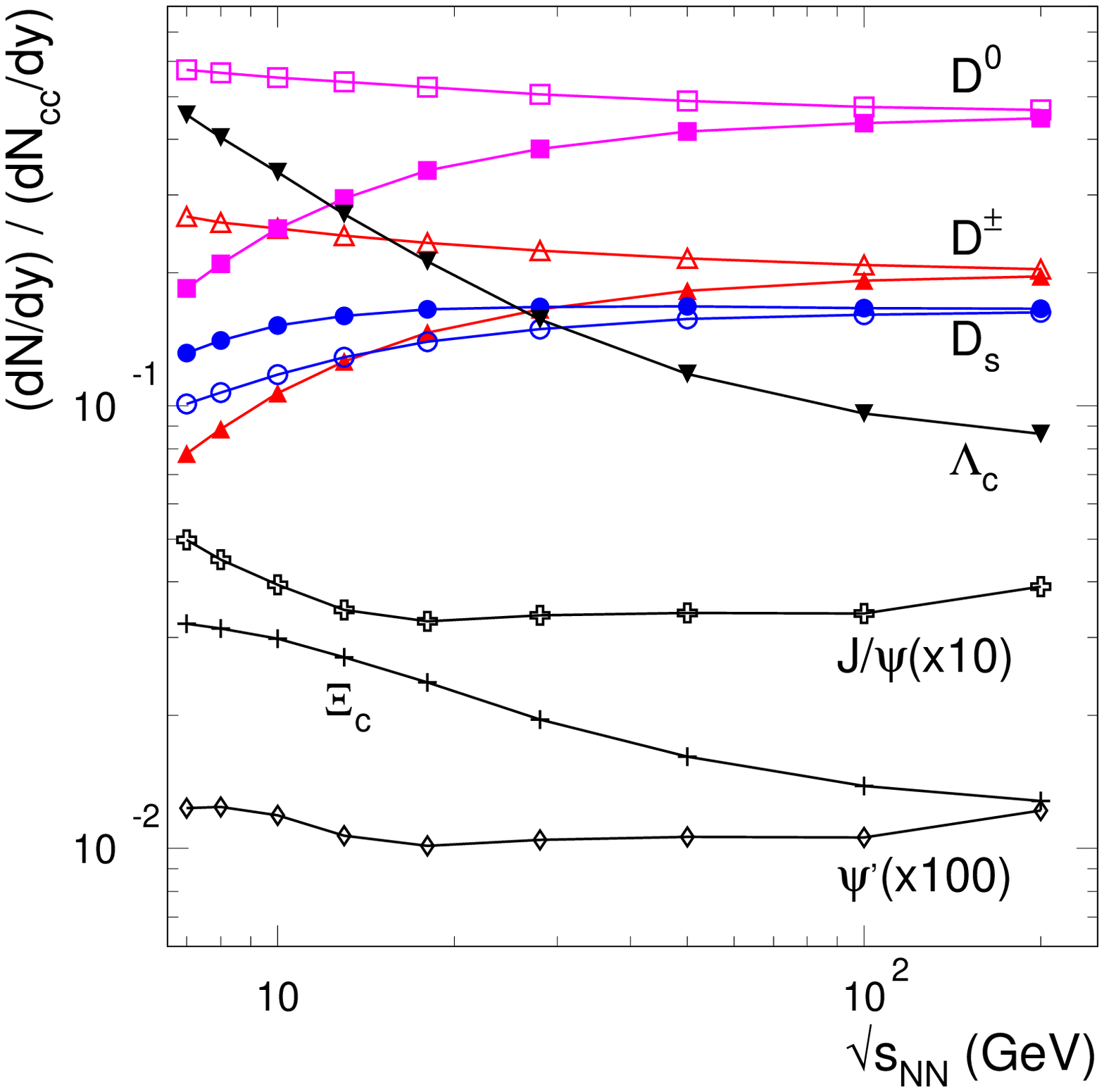}
\end{minipage}  & \begin{minipage}{.49\textwidth}
%charm#s opt=dm
\centering\includegraphics[width=.94\textwidth]{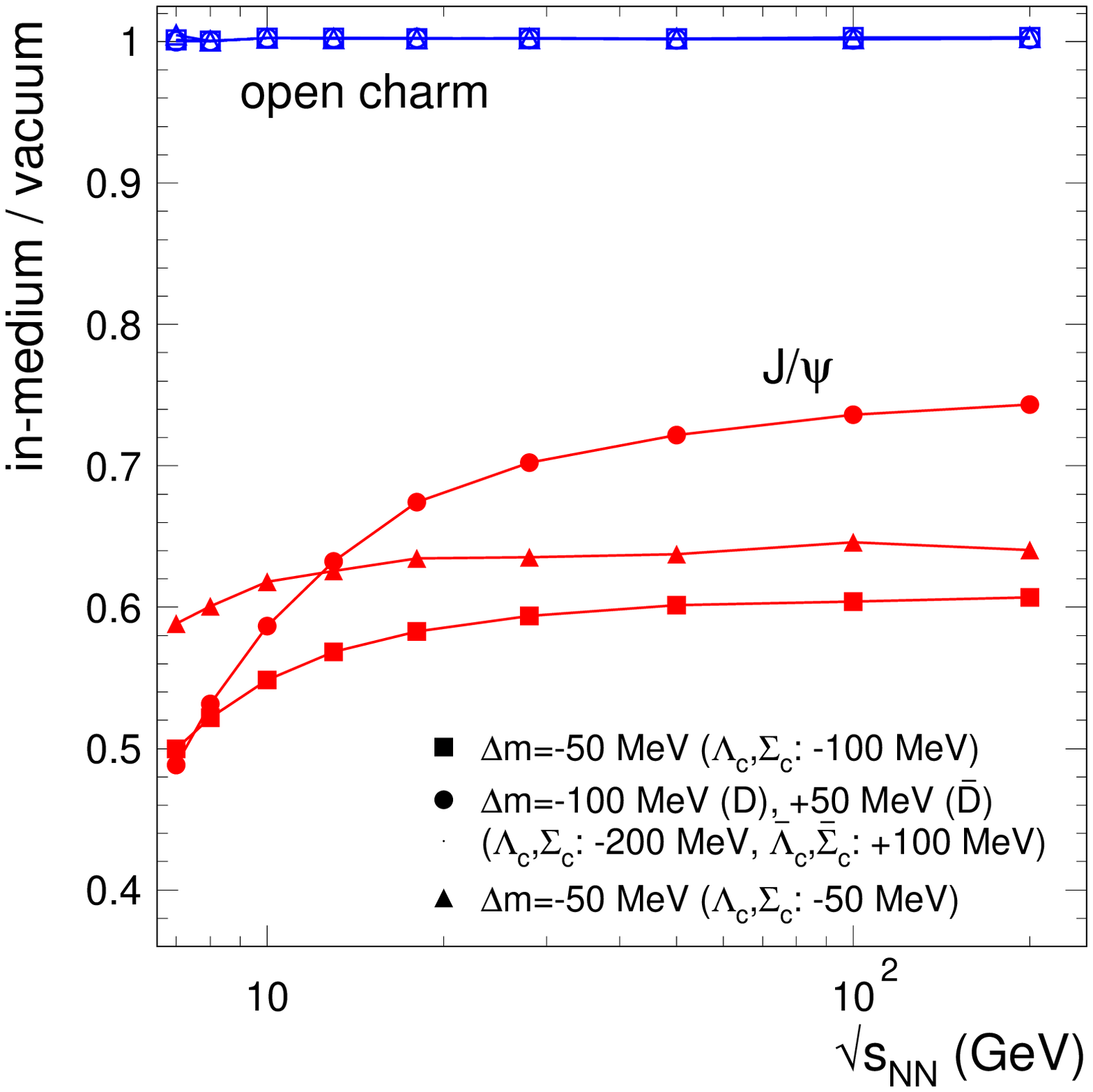}
\end{minipage}\end{tabular}
\vspace{-.5cm}
\caption{Left panel: energy dependence of the production yield relative to the
number of $c\bar{c}$ pairs for open and hidden charm hadrons 
(for mesons the open symbols are for the antiparticles).
Right panel: relative change in the production yield of all open charm 
hadrons and of $J/\psi$ meson considering different scenarios for 
in-medium mass modifications (see text).} 
\label{aa_fig3}
\end{figure}

A comprehensive set of model predictions from RHIC 
energy down to the charm production threshold is presented 
in Fig.~\ref{aa_fig3} for central Au-Au collisions ($N_{part}$=350).  
The left panel shows our predictions for the energy 
dependence of midrapidity yields (relative to the $c\bar{c}$ yield) for various
charmed hadrons.
The most striking behavior is observed for the production of 
$\Lambda_c^+$ baryons: their yield rises significantly towards lower energies.
In our approach this is caused by the increase in baryochemical potential 
towards lower energies (coupled with the charm neutrality condition). 
A similar behavior is seen for the $\Xi_c^+$ baryon.  
The relative production yields of D-mesons depend on their quark content 
and depend on energy only around threshold.
These results emphasize the importance of measuring, at low energies, 
in addition to D-mesons, also the yield of charmed baryons to get 
a complete measure of the total charm production cross section.

One of the motivations for the study of charm production at low energies
was the expectation \cite{cbm1,tol} to provide, by a measurement of D-meson 
production near threshold, information on their possible in-medium modification 
near the phase boundary. 
However, the cross section $\sigma_{c \bar c}$ is governed by the mass of 
the charm quark $m_c \approx 1.3$ GeV, which is much larger than any soft 
Quantum Chromodynamics (QCD) scale such as $\Lambda_{QCD}$. 
Therefore we expect no medium effects on this quantity.
The much later formed D-mesons, or other charmed hadrons, may well change
their mass in the hot medium. 
Whatever the medium effects may be, they can, because of the charm 
conservation, $\sigma_{c \bar c} = \frac{1}{2} ( \sigma_D +
\sigma_{\Lambda_c} +\sigma_{\Xi_c} + ...) + ( \sigma_{\eta_c} +
\sigma_{J/\psi} + \sigma_{\chi_c} + ...)$, in first order only lead to a 
redistribution of charm quarks \cite{aa2}.
This argument is essentially model-independent and applies equally 
at all energies.
This is demonstrated in the right panel of Fig.~\ref{aa_fig3}, where we plot
the relative change of the yields for different in-medium mass scenarios 
(see ref.~\cite{aa2} for details and references therein) compared to the case 
of vacuum masses.
In contrast, the yields of charmonia do vary and this is more prominent at
threshold energies (see Fig.~\ref{aa_fig3}).

%\section{Conclusions}

We have shown that the statistical hadronization model describes well the 
measured decrease with centrality and the rapidity dependence of $R_{AA}^{J/\psi}$ 
at RHIC energy. 
Extrapolation to LHC energy leads, contrary to the observations at RHIC, 
to $R_{AA}^{J/\psi}$ increasing with collision centrality. 
The increasing importance at lower energies of $\Lambda_c$ production was 
discussed and provides a challenge for future experiments. 
We have also shown that possible modifications of charmed hadrons 
in the hot hadronic medium do not lead to measurable changes in the
cross sections for production of hadrons with open charm. 
A possible influence of medium effects can be seen, however, in the 
yields of charmonium.


\section*{References}
\begin{thebibliography}{30}
\bibitem{satz} T. Matsui, H. Satz, Phys. Lett. B 178 (1986) 416.

\bibitem{satz2} F. Karsch, D. Kharzeev, H. Satz, Phys. Lett. B 637 (2006) 75.
%[hep-ph/0512239]; 
H. Satz, Nucl. Phys. A 783 (2007) 249. % [hep-ph/0609197].

\bibitem{aa2} A. Andronic, P. Braun-Munzinger, K. Redlich, J. Stachel, 
Nucl. Phys. A 789 (2007) 334; % [nucl-th/0611023].
%\bibitem{aa3} A. Andronic, P. Braun-Munzinger, K. Redlich, J. Stachel, 
Phys. Lett. B 652 (2007) 259; % [nucl-th/0701079].
%\bibitem{aa4} A. Andronic, P. Braun-Munzinger, K. Redlich, J. Stachel, 
Phys. Lett. B 659 (2008) 149. % [arXiv:0708.1488].

\bibitem{pbm1} P. Braun-Munzinger, J. Stachel,
Phys. Lett. B 490 (2000) 196; % [nucl-th/0007059];
Nucl. Phys. A 690 (2001) 119c.% [nucl-th/0012064].

\bibitem{aa1} A. Andronic, P. Braun-Munzinger, K. Redlich, J. Stachel, 
Phys. Lett. B 571 (2003) 36. % [nucl-th/0303036].

\bibitem{cac} 
M. Cacciari, P. Nason, R. Vogt, Phys. Rev. Lett. 95 (2005) 
122001. % [hep-ph/0502203].

\bibitem{rv1} ALICE Collaboration, J. Phys. G 32 (2006) 1295;
R. Vogt, Int. J. Mod. Phys. E 12 (2003) 211. % [hep-ph/0111271].

\bibitem{cc1} A. Adare et al. (PHENIX), Phys. Rev. Lett. 97 (2006) 252002.
%[hep-ex/0609010].

\bibitem{aat} A. Andronic, P. Braun-Munzinger, J. Stachel, 
Nucl. Phys. A 772 (2006) 167, % [nucl-th/0511071].

\bibitem{phe1} A. Adare et al. (PHENIX),  Phys. Rev. Lett. 98 (2007) 232301.
%[nucl-ex/0611020].

\bibitem{phe2} A. Adare et al. (PHENIX), arXiv:0711.3917.

\bibitem{cbm1} P. Senger, J. Phys. Conf. Series 50 (2006) 357.
%\bibitem{aa5} A. Andronic, P. Braun-Munzinger, K. Redlich, J. Stachel, 
%arXiv:0707.4075 
\bibitem{tol} L. Tolos,  J. Schaffner-Bielich, H. St\"ocker,
Phys. Lett. B 635 (2006) 85. % [nucl-th/0509054].

\end{thebibliography}
\end{document}